\documentclass[12pt,preprint]{aastex}

\begin{document}

\title{THE NATURE OF THE GOULD BELT FROM A \\
FRACTAL ANALYSIS OF ITS STELLAR POPULATION}

\author{N\'estor S\'anchez,\altaffilmark{1}
        Emilio J. Alfaro,\altaffilmark{1}
        Federico Elias,\altaffilmark{2} \\
        Antonio J. Delgado,\altaffilmark{1} and
        Jes\'us Cabrera-Ca\~no\altaffilmark{3}}

\altaffiltext{1}{Instituto de Astrof\'{\i}sica de Andaluc\'{\i}a,
                 CSIC, Apdo. 3004, E-18080, Granada, Spain;
                 nestor@iaa.es, emilio@iaa.es, delgado@iaa.es.}
\altaffiltext{2}{Instituto de Astronom\'{\i}a, Universidad Nacional
                 Aut\'onoma de M\'exico, C. P. 04510, M\'exico D. F.,
                 M\'exico; felias@astroscu.unam.mx.}
\altaffiltext{3}{Facultad de F\'{\i}sica, Departamento de F\'{\i}sica
                 At\'omica, Molecular, y Nuclear, Universidad de Sevilla,
                 Apartado 1065, 41080 Sevilla, Spain; jcc-famn@us.es.}

\slugcomment{The Astrophysical Journal: accepted}

\begin{abstract}
The Gould Belt (GB) is a system of gas and young, bright stars
distributed along a plane that is inclined with respect to the
main plane of the Milky Way. Observational evidence suggests
that the GB is our closest star formation complex, but its true
nature and origin remain rather controversial. In this work we
analyze the fractal structure of the stellar component of the
GB. In order to do this, we tailor and apply an algorithm that
estimates the fractal dimension in a precise and accurate way,
avoiding both boundary and small data set problems. We find that
early OB stars (of spectral types earlier than B4) in the GB have
a fractal dimension very similar to that of the gas clouds in our
Galaxy. On the contrary, stars in the GB of later spectral types
show a larger fractal dimension, similar to that found for OB
stars of both age groups in the local Galactic disk (LGD). This
result seems to indicate that while the younger OB stars in the
GB preserve the memory of the spatial structure of the cloud
where they were born, older stars are distributed following a
similar morphology as that found for the LGD stars. The possible
causes for these differences are discussed.
\end{abstract}


\section{INTRODUCTION}

The brightest stars near the Sun are mainly distributed along two
great circles over the sky: the Milky Way plane and another strip
of bright stars known as the Gould Belt (GB) which is inclined
respect to the Galactic plane \citep{Gou79,Sto74}. Although the
existence of the GB was formerly reported as soon as in 1847 by
Sir John F. W. \citet{Her47} from naked eye observations
of the southern sky, its nature and origin are still poorly
understood. The GB is a complex system of gas and stars composed
not only by single massive stars but also by OB associations and
an interstellar medium (dust, neutral hydrogen, and molecular
clouds) that shows some kinetic features which have been considered
by several authors as connected to the stellar population \citep{Lin73}.
The GB was firstly interpreted as an apparently expanding ring (or
torus) of young massive stars, but later, with the detection of
young low-mass stars, its structure seemed to be better represented
by an inhomogeneous disk \citep{Gui98}. Recent studies characterize
it as a stellar disk having elliptical shape, with a major semiaxis
of $\sim 600$ pc and minor semiaxis of $\sim 400$ pc, which is inclined
around 18$^{\circ}$ to the Galactic plane \citep{Eli06a}. The stellar
component exhibits a range of ages lesser than $100$ million years
\citep{Tor00}. Given its size ($\sim 1$ kpc), age range, and
relationship with the gas in different phases, it has been
suggested that the GB would be our closest star formation complex
\citep{Elm00}, in the sense given to this concept by \citet{Efr95},
i.e., the largest region of a galaxy showing a coherent star
formation process, where the term ``coherent" should be
interpreted as originated from the same, monoparental,
giant gas cloud.

However, there exist both theoretical and observational results
showing inconsistencies with this scenario: the age pattern of
OB associations is irregular, with the youngest ones located well
outside the expanding gas ring \citep{Per03}; the kinematics of
the gas does not seem to be in agreement with the velocity field
of the stellar component; and the modelization of gas dynamics is
not compatible with the estimated radius of the stellar component,
as well as with its range of ages \citep{Per03,Mor99}. Moreover, the
analysis of the stellar spatial velocity diagram in the $U-V$ plane
shows a clear bimodal distribution \citep{Eli06b}.
All these results make it difficult to imagine
a single monoparental origin for the GB and
suggest a GB formed by a spurious concatenation of different
stellar subsystems whose formation processes were not necessarily
connected among them. This point is particularly important for a
full understanding of the mechanisms involved in the formation of
the Milky Way. Can we design some kind of experiment able to
answer this question? Here we propose to analyze the fractal
structure of the stellar component of the GB in order to go
deeper into its nature and origin. The base of this approach
lies in the fact that the interstellar medium seems to show
a fractal structure when observed at the scale of clouds
\citep[][and references therein]{Ber07} and a
multi-scaling behavior \citep{Cha01} when galactic scales are
considered. Thus, stars forming from the same cloud should exhibit
fractal patterns too if their birth places uniformly follow the
densest regions \citep{Elm01}. In this work, we first develop a
``reliable" algorithm to compute fractal dimensions of a sample
of discrete points (\S~\ref{sec_algorithm}) and then we
use it to study the distribution of stars in the GB
(\S~\ref{sec_gould}). The main conclusions are
summarized in \S~\ref{sec_conclusions}.

\section{ESTIMATING THE FRACTAL DIMENSION}
\label{sec_algorithm}

Strictly speaking, a fractal is defined as an object whose
Hausdorff dimension is larger than its topological dimension
\citep{Man83}. To estimate the Hausdorff dimension, authors
use different working definitions that fit their methods and
needs, and thus there is not a unique definition of fractal
dimension. When dealing with a distribution of points in space
it is very useful the so-called ``correlation dimension"
\citep{Gra83}. This is a widely used method because of its
robustness and because it is relatively easy to implement
on real (experimental or observational) data.

Let us consider a distribution of $N$ points in space with
positions ${\bf x}$. The number of other points within a
sphere of radius $r$ centered on the $i$-th point is given
by the expression:
\begin{equation}
\label{eqnr}
n_i(r) = \sum_{j=1,\ j \neq i}^{N}
H \left( r- | {\bf x}_i-{\bf x}_j | \right) \ \ ,
\end{equation}
where $H(x)$ is the Heaviside step function. This number can
be evaluated by choosing $M$ different points as centers and
then averaging to obtain the probability of finding a point
within a sphere of radius $r$ centered on another point. This
probability is expressed in the form
\begin{equation}
\label{eqcr}
C(r) = \frac{1}{M(N-1)} \sum_{i=1}^{M} n_i(r) \ \ .
\end{equation}
For a fractal set this quantity, called the correlation
integral, scales at small $r$ as
\begin{equation}
\label{eqdc}
C(r) \sim r^{D_c} \ \ ,
\end{equation}
being $D_c$ the correlation dimension, which in practice
can be identified with the slope of the best fit in a
$\log C(r) - \log r$ plot. In other words, the correlation
dimension tells us how it changes (as $r$ increases) the
probability that two points chosen randomly are separated
by a distance smaller than $r$. For a homogeneous distribution
of points in space we expect $D_c = 3$, whereas in a plane
$D_c = 2$. If points are distributed obeying a fractal
geometry then $D_c < 3$ (in 3D-space) or $D_c < 2$ (in a
plane).

\begin{figure}
\plotone{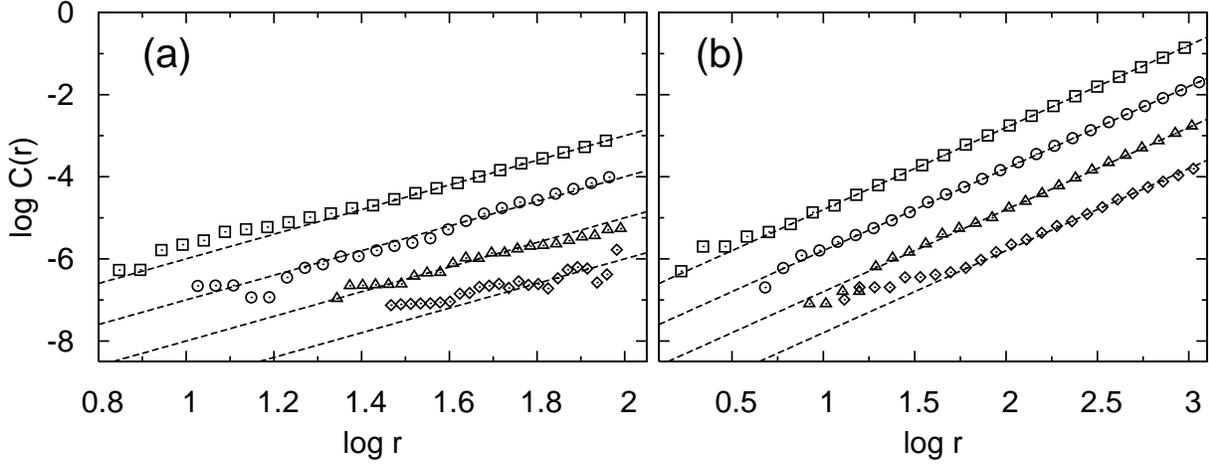}
\caption{Correlation integral $C(r)$ for points distributed randomly
within a disk of radius $R_d = 2500$ pc and half-heigh $Z_{max}
= 100$ pc. The number of points is $N=2000$ (squares), $1000$
(circles), $500$ (triangles), and $200$ (rhombuses). All the
curves have been arbitrarily shifted downward (except the upper
one) for clarity. {\bf (a)} The calculation is done in the
three-dimensional space using spheres with radius ranging from
the minimal distance between two points to a maximum of $r_{max}
= Z_{max}$. As a reference, the dashed lines show the expected
slopes of 3 for these cases. For the case $N=200$ points, it is
difficult to infer the expected linear behavior. Panel {\bf (b)}
shows the results when the calculation is done over the distributions
projected on the $Z=0$ plane and using circles with different radii
$r$ up to a maximum of $r_{max}=R_d/2$. In these cases the expected
value of the slopes (dashed lines) is 2. The linear behaviors are
clearly observed even for a relatively low number of points.}
\label{teoria}
\end{figure}

Typically, when evaluating $D_c$ for real data (and not for
infinite, perfect fractals) the power-law scaling relation
(eq.~[\ref{eqdc}]) is valid only within a limited range of $r$
values \citep{Smi88}. At very small scales (of the order of
the mean distance to the nearest neighbor) the distribution
looks like a set of isolated points and $D_c$ tends to zero.
On the other hand, the finite size of the set also results
in decreased $D_c$ values at large $r$ values (of the order
of the set size), because near the edge each point is surrounded
by other points only on one side and then $C(r)$ tends to be
underestimated. It has been proved that these and many other
effects can lead to bad estimations of $D_c$, mainly when too
few data are available \citep{Smi88,Kit00,Cic02}. In order to
assess how the limited number of data points would alter our
estimation of $D_c$ for the GB, we have done some numerical
experiments. Figure~\ref{teoria} shows $\log(C)-\log(r)$ plots
for random distributions of points within disk-like structures.
Boundary effects were avoided by keeping the condition that
sampling spheres always are inside the volume occupied by the
disk. For the three-dimensional case (Fig.~\ref{teoria}a), the
expected linear behavior is observed at high $r$ values for a
relatively high number of points $N$, but at low $r$ values we
see departures from the linear behavior arising from the lack
of statistic of finite samples \citep{Smi88}. This tendence
becomes more evident as $N$ decreases because
the mean distance between nearest neighbor
increases. For the case $N=200$ points it is very difficult to
infer a linear behavior at all. This is an important problem
because many times the number of available data points is rather
small. In these cases the estimated fractal dimension depends
strongly on the spatial range used to calculate the slope in the
$\log(C)-\log(r)$ plot \citep{Pis95}. However, when the calculation
is done on the projected distributions (Fig.~\ref{teoria}b), the
range of $r$ values used to estimate the integral correlation can
be extended to higher values, while still keeping the spheres
(that actually become circles in the 2D projection) inside the
sample. In this case, the linear behavior is clearly appreciated
-at relatively high $r$ values-  even when only $N=200$ points
are used. The average $D_c$ values for ten random realizations,
calculated by doing linear fits in the range $\log r \geq 1.5$,
are shown in Table~\ref{table1}. The results are always close to the
theoretical values (3 or 2) but the standard deviations become
quite high in the 3D case for a relatively low number of points.
In any case, the ``problematic" region of relatively low $r$ values
has to be excluded of the linear fit. In order to do this, we impose
the condition that the standard deviation of each $C(r)$ value must
be smaller than the corresponding $C(r)$ value. This simple criterion
eliminates poorly estimated data (i.e., bad sampling) ocurring mainly
at too low $r$ values.

\begin{figure}
\plotone{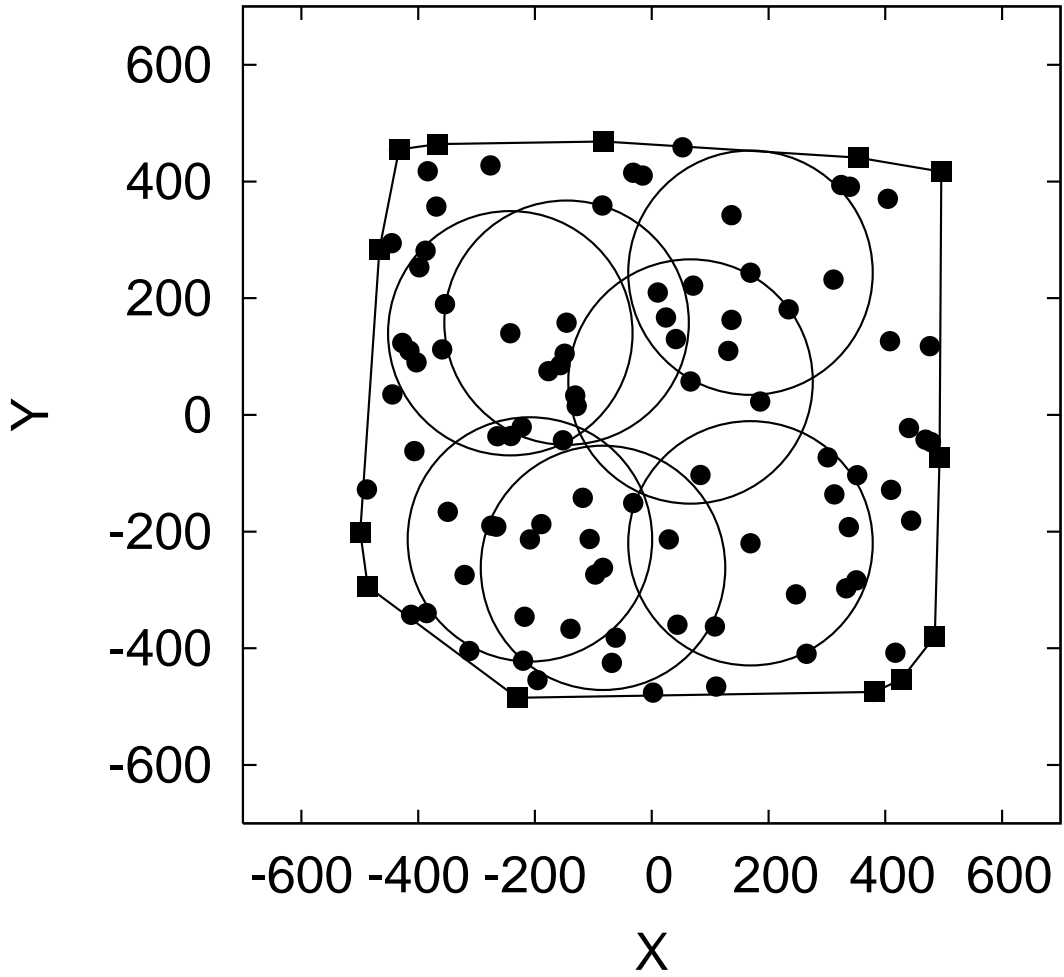}
\caption{Example of random distribution of 100 points in a square
region of side 1000. The vertices of the convex hull (see text) are
indicated by squares connected by lines. The circles (with radius
200) illustrate how the sampling is done by keeping them inside the
boundary defined by the convex hull.}
\label{convex}
\end{figure}

\begin{figure}
\plotone{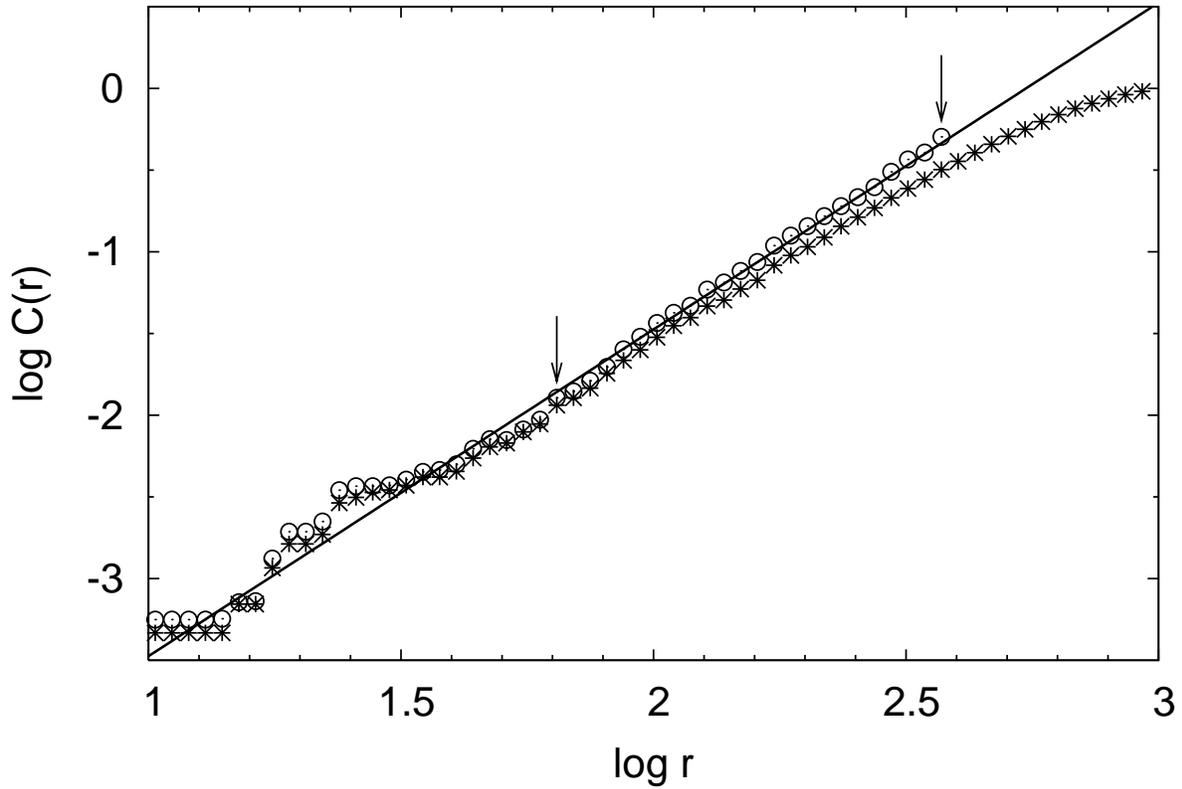}
\caption{Correlation integral $C(r)$ for the same point set shown
in Figure~\ref{convex}, when the calculations are performed
by taking (open circles) and not taking (asterisks) into
account boundary effects. The solid line indicates the expected
slope for a random distribution of points ($D_c=2$), and the
vertical arrows indicate the range for which the standard
deviation of $C(r)$ is smaller than $C(r)$.}
\label{ejemplo}
\end{figure}

As an example, Figure~\ref{convex} shows a random distribution
of 100 points in a square region of side 1000. In order to take into
account edge effects, we first determine which points of the sample
are vertices of the convex hull, i.e. we determine the minimum-area
convex polygon containing the whole set of data points
(square symbols connected by lines in Figure~\ref{convex}).
For this we use the algorithm proposed by \citet{Edd77}.
Now we can place circles of different radii $r$ to
evaluate $C(r)$ according to equation~(\ref{eqcr}).
The result is plotted in Figure~\ref{ejemplo}. The calculations
were carried out both taking (open circles) and not taking
(asterisks) into account boundary effects. If circles are
allowed to cross boundaries $C(r)$ tends to be underestimated,
and this effects is higher the higher the radius $r$. For this
reason $D_c$ would tend to be smaller than the expected value
if this effect is not considered (for comparison, solid line
in this figure shows the expected $D_c=2$ result). The range
of $r$ values that fulfill the condition that the standard
deviation of $C(r)$ is smaller than $C(r)$ is indicated in
Figure~\ref{ejemplo} as vertical arrows. For $r$ values
below this range the departures from the linear behavior
become more evident.

\begin{figure}
\plotone{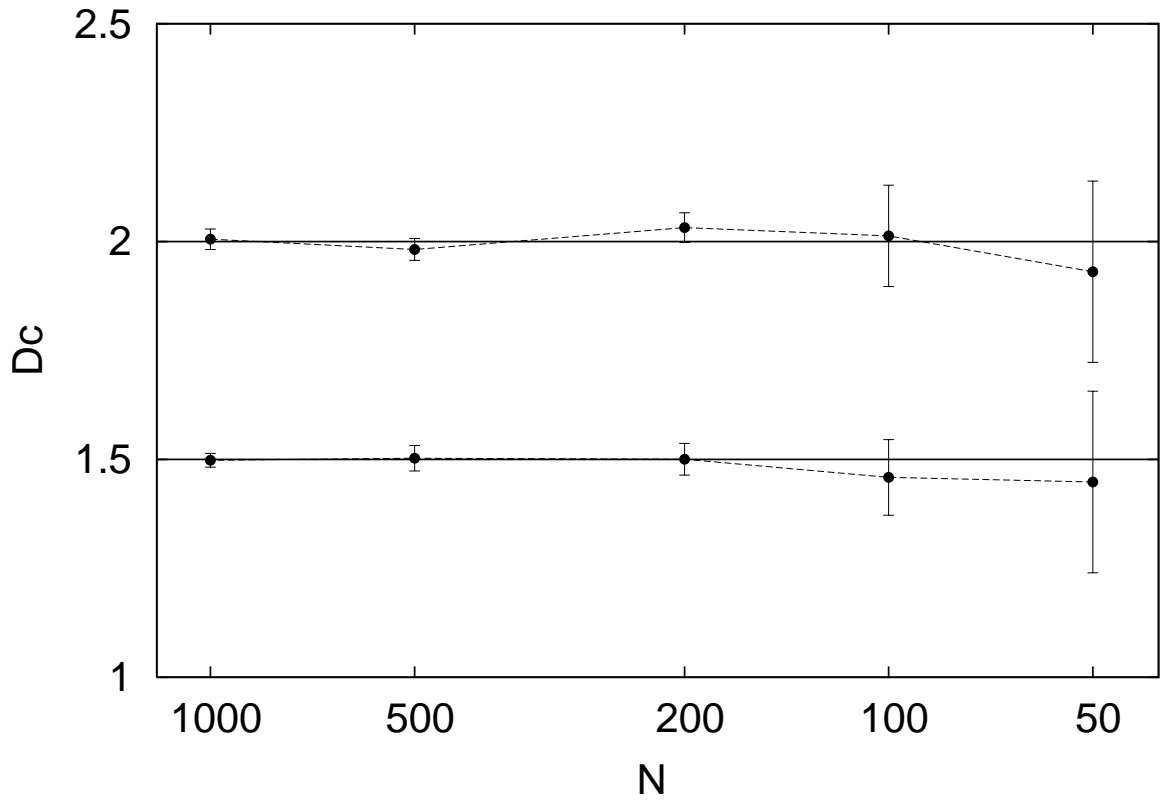}
\caption{Correlation dimension $D_c$ as a function of the
sample size $N$ for fractal distributions of points.
Each result is the average of 10 different realizations,
and the bars are the average of the uncertinties calculated
using bootstrapping (see text). The horizontal lines
indicate the fractal dimension used to generate the
distribution of points (1.5 and 2).}
\label{prueba}
\end{figure}

The algorithm we have developed follows the next steps: first, the
three-dimensional distribution of points is projected on its mean
plane; the boundary is determined finding the convex hull
of the sample; the correlation $C(r)$ is calculated using
equations~(\ref{eqnr}) and (\ref{eqcr}) but always keeping the circles
inside the sample boundary; and then the correlation dimension
is determined as the slope of the best $\log C(r) - \log r$
linear fit. Very low $r$ values (and consequently poorly
estimated $C(r)$ values) are excluded from the fit. Finally,
we use bootstrap techniques to estimate the uncertainty of the
calculated value: we repeat the calculation of $D_c$ on a series
of random resamplings of the data and the standard deviation of
the obtained set of fractal dimensions is taken as the error in
our estimation. When working with relatively thin disks the
situation is practically the same as taking a two-dimensional
slice of a spherical, three-dimensional distribution of points
within a sphere. In this case, the fractal dimension of the
three-dimensional distribution of points $D_c({\rm 3D})$ and
the two-dimensional dimension $D_c({\rm 2D})$ are related
through the simple expresion \citep{Fal90}:
\begin{equation}
D_c({\rm 2D}) = D_c({\rm 3D}) - 1 \ \ .
\end{equation}
We have simulated two-dimensional fractal distributions of points
in order to test the algorithm performance. The fractals were
simulated by placing four squares of radius $R/L$ (with $L \geq
2$) inside a square of radius $R$ (we place one square in each
quadrant). The procedure is repeated successively 10 times to
obtain $4^{10}$ ($\sim 10^6$) points distributed according
to a fractal pattern with dimension $D_c = \log 4 / \log L$.
Finally, we randomly removed points from the fractal until
reaching a given sample size $N$. Figure~\ref{prueba} shows
some example results for the cases $D_c = 1.5$ ($L \simeq
2.5$) and $D_c = 2$ ($L = 2$). Even for relatively small
sample sizes ($N \sim 100$) the algorithm is able to estimate
$D_c$ in a reliable way.

\section{THE FRACTAL DIMENSION OF THE GOULD BELT}
\label{sec_gould}

\begin{figure}
\plotone{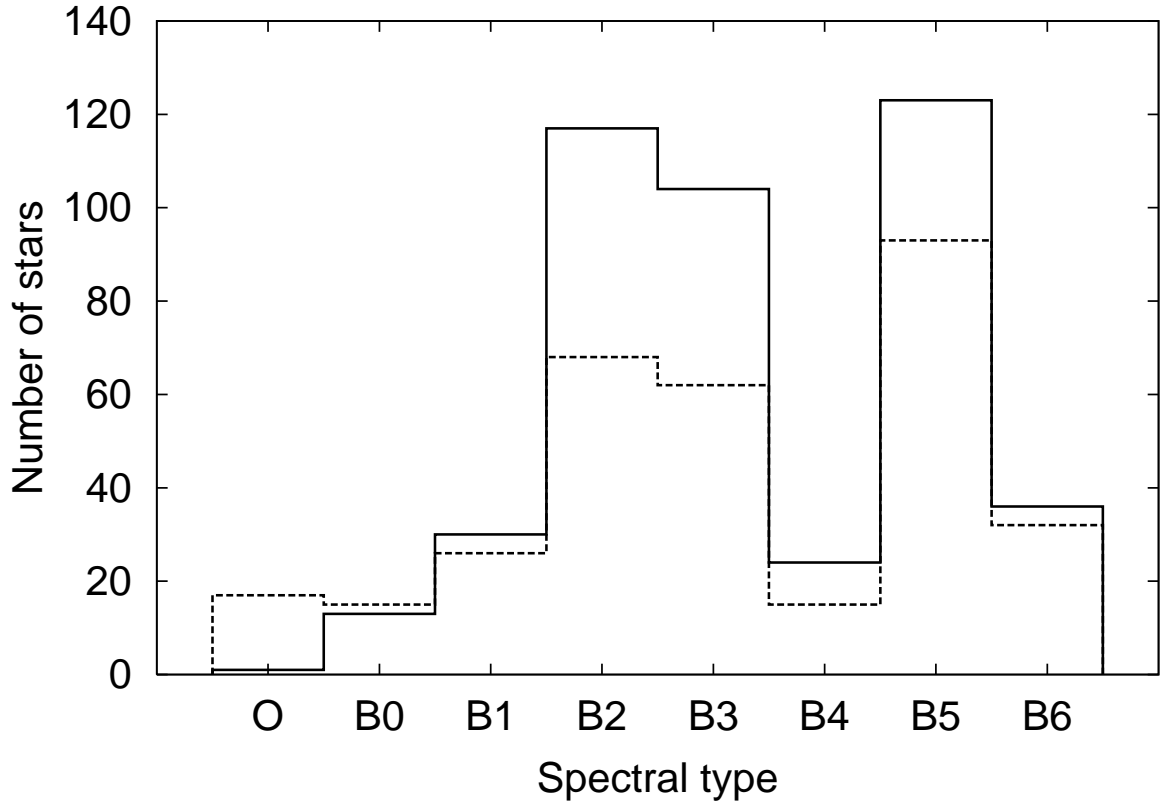}
\caption{Distribution of spectral types for the sample
of stars taken from \citet{Eli06a}. Continuous line
refers to stars belonging to the GB, whereas dashed
line refers to the LGD.}
\label{mktipos}
\end{figure}

\begin{figure}
\plotone{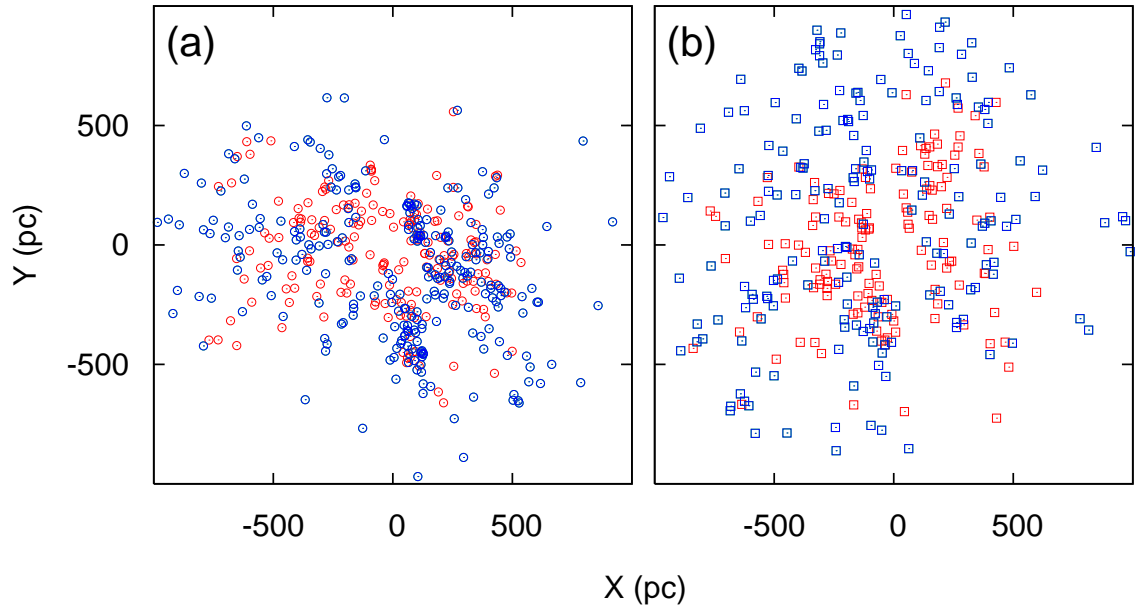}
\caption{Positions of the stars in our sample. Stars
are represented as circles (for the GB) or squares (for
the LGD). The blue symbols refer to the spectral range O-B3 and
red symbols to the range B4-B6. The panels show the projections
on the mean planes for each distribution, both for the GB (a)
and for the LGD (b).}
\label{mapas}
\end{figure}

\begin{figure}
\plotone{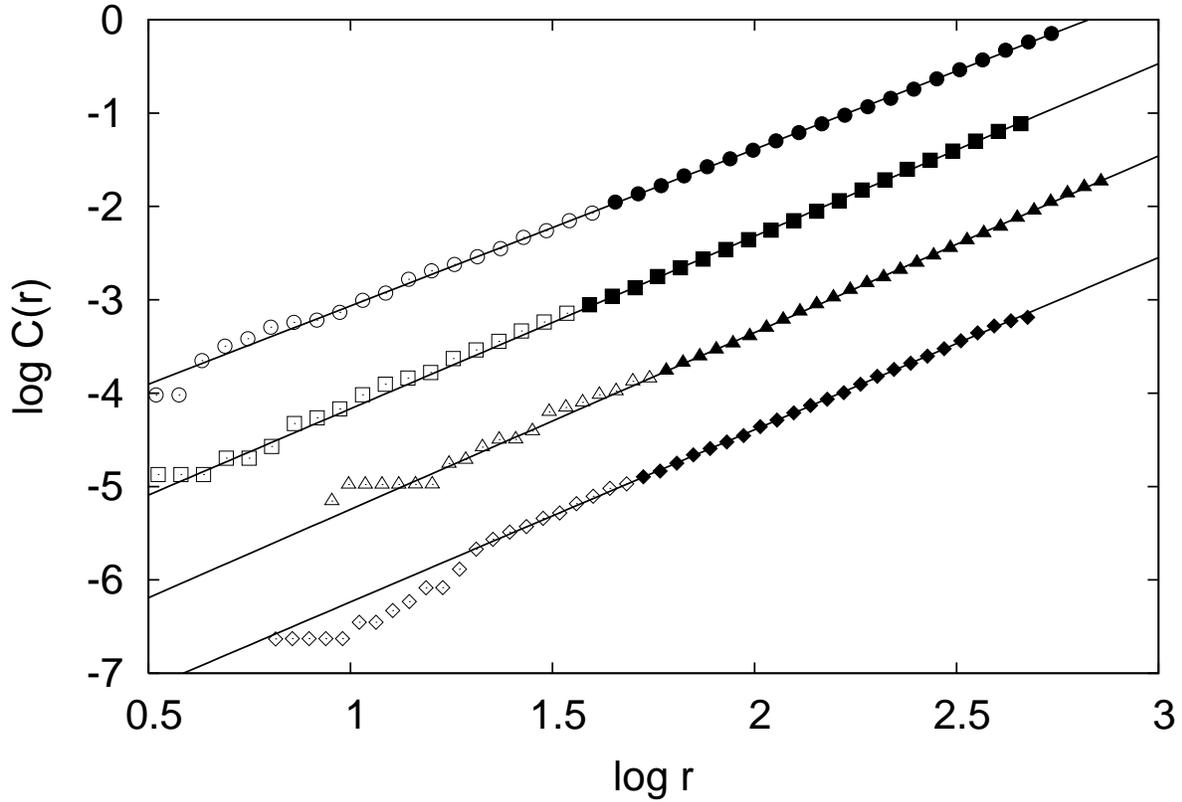}
\caption{Projected correlation integral for stars
in the solar neighborhood. The calculations were done
for GB early stars (circles), GB late stars (squares), LGD
early stars (triangles) and LGD late stars (rhombuses).
Filled symbols indicate the range used for the linear
fits (shown as lines). All the curves have been arbitrarily
shifted downward (except the upper one) for clarity.}
\label{correla}
\end{figure}

The first problem that arises when studying the Gould Belt
is how to select the stars belonging to this system.
\citet{Eli06a} developed a new method to perform a
three-dimensional spatial classification that was applied
to a sample of 553 OB stars from the {\it Hipparcos} catalog
with precise distances of less than 1 kpc.
This allowed them to separate and estimate the spatial
structure of the stars belonging to the GB and to the LGD.
The distributions of MK spectral types for the sample of
stars used \citep{Eli06a} are shown in Figure~\ref{mktipos}.
We see two populations clearly differentiated, one with
spectral types earlier than B4 (which we will call ``early"
stars) and another one with spectral types B4-B6 (``late"
stars), that can be associated with two age groups centered
at $\sim 20$ Myr and $\sim 70$ Myr, respectively.
The spatial distribution of these stars is shown
in Figure~\ref{mapas}. The projections on each of the mean
planes clearly reveal a clumpy, filamentary structure for
the GB (Fig.~\ref{mapas}a), whereas the stars in the LGD
seem to be distributed more homogeneously (Fig.~\ref{mapas}b).
The almost free of stars gap seen in Figure~\ref{mapas}
corresponds to the line of nodes in which both the GB and
the LGD coexist. Close to this line, the probability of
belonging to the GB has a high uncertainty because the
bayesian probability in this region depends only on the
``a priori" probability (i.e., the relative frequencies
of both systems) and not on the probability conditioned
to the spatial position \citep[see][]{Eli06a}. This gap
is almost imperceptible for the GB and notorious for the
the LGD simply because in the sample of stars used there
are more stars in the GB than in the LGD.
What we have done in this work is to quantify the degree of
inhomogeneity by calculating the fractal dimension\footnote{The
term ``fractal dimension" alone may be ambiguous because
there exists several different definitions of this quantity.
We will indifferently use the term ``fractal dimension" or
``correlation dimension" to refer to $D_c$.} for the GB and
LGD (both early and late stars). The results are
shown in Figure~\ref{correla}. For all cases the behavior
is almost perfectly linear in this log-log plot, with correlation
coefficients $\simeq 0.99$. The number of circles used to evaluate
$C(r)$ for each $r$ (i.e. $M$ in eq.~[\ref{eqcr}]) is usually of the
order of $10^2$. However, strictly speaking, $M$ depends on $r$ in
the sense that larger $r$ allows fewer circles within the borders.
For the largest $r$ values considered $M$ is of the order of $10$.
The slopes of the linear fits give the following fractal dimension
values:\\ \\
\begin{tabular}{ll}
GB-early: & $D_c({\rm 3D})=2.68\pm 0.04$, \\
GB-late:  & $D_c({\rm 3D})=2.85\pm 0.04$, \\
LGD-early:& $D_c({\rm 3D})=2.89\pm 0.06$, \\
LGD-late: & $D_c({\rm 3D})=2.84\pm 0.06$.
\end{tabular}\\
\ \\
If the empty gap mentioned before were producing some bias in
the determination of $D_c$ for the LGD then the unbiased values
should be even higher than the obtained ones.

We see that the distribution of stars in the solar neighborhood
exhibits a certain degree of fractality, with $2.7 \lesssim D_f
\lesssim 2.9$. Statistical tests show that there is a difference
between early stars in the GB and the other subsets that is, in
the worst of the cases, statistically significant. Therefore,
interestingly, the fractal dimension of early stars in the GB
($D_f \sim 2.7$) is significantly smaller than that of the rest
of the sample. Thus, these early stars have similar ages and are
distributed following fractal patterns analogous to those observed
in the gas of star forming regions in the Galaxy \citep{San05,San07}.
Therefore, it seems very likely that this group of stars was born
from the same parent cloud. In contrast, later stars in the GB have
a somewhat different, more homogeneous type of distribution with a
fractal dimension similar to that obtained for the LGD ($D_f \gtrsim
2.85$).

What is the origin of this difference? First, we have to point out
that, in principle, different global properties are expected for the
interstellar medium (ISM) at different spatial and/or temporal scales
because the main physical mechanisms acting at each scale are not
necessarily the same. A monolithic gas cloud can be characterized
as a turbulent medium, and this turbulence can be driven by many
energy sources \citep{Elm04}. However, at larger spatial scales, the
ISM in a spiral arm is influenced by other disturbances (gravitational,
magnetic, etc.) that modify the gas distribution. The fractal analysis
gives us a simple but objective measurement of such structure through
the degree of smoothness or clumpyness. This kind of analysis has been
applied extensively to evaluate the structure of interstellar gas from
parsec to kiloparsec scales, as well as the distribution of stars, star
clusters, and star forming sites [see
\citet{Car06,Kha04,Wes99,Elm06,Car04,Fue06a,Fue06b,Fue06c,Ode06}
for some recent examples]. But often the analysis techniques are so
diverse that it is not quite easy to extract robust conclusions. It
seems that the average fractal dimension of single gas clouds is
around $D_f \sim 2.7$ \citep{San05,San07}, although it may vary
when the spatial scale is increased and when the distribution of
gas clouds in the galactic disk is analyzed \citep{Cha01}. But what
about the stars? Young, new-born stars will reflect the conditions
of the ISM from which they were formed. Therefore, a group of stars
born from the same monolithic cloud, i.e. born at almost the same place
and time, should have a fractal dimension similar to that of the parent
cloud. This is exactly what we find for early-type stars in the GB: we
obtain a fractal dimension very similar to the ISM value. Otherwise, if
the star sample is representative of a population born from various clouds
and/or with different star formation histories, then the fractal dimension
will represent the gas distribution at different spatial or temporal
scales according to a multi-scale structure of the ISM \citep{Cha01}.
At galactic level, the fractal dimension of the distribution of stars
and/or star forming regions exhibits a very wide range of values, but
until now no correlation has been clearly found between fractal patterns
at galactic level and other galactic properties \citep{Ode06,Fei87,Par03}.
There are, however, some suggestions that the fractal dimension of the
distribution of stars and star forming sites increases with time after
the star formation process \citep{Fue06a,Fue06b,Fue06c,Ode06,Sch06},
probably due to the action of some physical mechanisms which tend to
reorganize/destroy the original structure.

Here we have found quite different values for the fractal dimension
of early-type stars in the GB and the LGD. This difference reflects
the fact that most of OB associations in the solar neighborhood are
located in the GB \citep{Zee99,Eli06a}. This is, the distribution
of OB stars in the LGD does not seem to show any kind of ``typical"
stellar grouping, whereas young OB stars in the GB exhibit a more
hierarchized distribution with multiple OB associations. In this
sense, we can say that the GB shows clear signatures of the internal
structure of its parental cloud. This clustering in the young
population of the GB is quantified through the fractal dimension,
whose value is very close to the value characterizing gas clouds
in the Galaxy. Thus, the possible causes for
different values for the fractal dimension of early-type stars in
the GB and the LGD can be studied by analyzing the possible causes
for different number of associations in both stellar systems. There
are three possible factors (or a combination of them) that can
contribute to this difference. Firstly, the age difference between
both groups could be significant enough to explain this result. It
is clear that the separation according to the spectral type of the
stars provides only a gross age separation. The spectral types allow
to derive an upper limit for the age but they do not give the real
lifetime of a star that also depends on the initial chemical
composition. Although the earlier spectral groups of both the
GB and the LGD have similar mean values and variances (see
Figure~\ref{mktipos}), there exists the possibility that one
group is actually younger than the other one. This age difference
could be enough to allow the LGD group to disperse showing a less
structured distribution, whereas the GB group would still trace the
density peaks of the parental cloud. Secondly, the difference in
the degree of clustering among both groups can be produced by
different physical conditions of the parental clouds. Much of
stellar dispersal from active star-forming regions occurs on
timescales of $10^6 - 10^7$ years, and there are several possible
scenarios to account for this dispersal of young stars from their
birth places \citep{Mam01}. The time scale for dispersal depends on
the range of values of the density peaks inside the cloud which, at
the same time, should depend on the local ISM pressure \citep{Elm06pre}.
Thus, clouds having different dynamical evolutions in different
environmental conditions would likely give rise to stellar groups
with different lifetimes. Depending on these critical values and
on the age range of the stellar sample, we can get either a complex of
associations or a uniform distribution of stars. Thirdly, different gas
distributions in the parental clouds can be responsible for variations
in the fractal dimension among young stars in the GB and the LGD. For
the GB, $D_f$ agrees well with the range of values expected for the
case of single monoparental clouds in the Galaxy, i.e., $D_f \simeq
2.5-2.7$ \citep{San05,San07}. On the other hand, the fact that the
fractal dimension of the LGD is higher than the expected for
individual clouds suggests that this population did not have a
monoparental origin. Even though the young stars in the LGD were
born in individual clouds each one having {\it internal} fractal
dimensions around $\sim 2.6$, the {\it set} of clouds that formed
this group were distributed following spatial pattern with a
higher fractal dimension at a larger scale according to a
multi-scale scenario \citep{Cha01}. In other words, even though
the two young stellar groups in the GB and the LGD were born at
the same time, the structure of the parental gas at the birth
time has likely driven the final geometric structure of these
stellar populations.

In order to restrict even more the problem under consideration,
we have selected subsamples of stars with spectral types earlier
than B2 both for the GB and for the LGD. Stars belonging to these
groups have ages younger than $\sim 10$ Myr. The determination of
the fractal dimension for these ``very early" subsamples yields
similar results: $D_c({\rm 3D})=2.62\pm 0.08$ for the GB and
$D_c({\rm 3D})=2.84\pm 0.08$ for the LGD. The uncertainties are
higher than before because of the smaller number of stars in the
samples.  Obviously this result does not rule out the possibility
that both groups have different ages but, in this case, it is more
difficult to interpret these results in terms of age differences
among groups.

With the available data we cannot unambiguously discriminate the most
likely explanation for the different fractal dimensions found for the
very early-type stars in the GB and the LGD. Maybe all of these
factors are contributing in some way to the problem. However, we
consider that the main factors are probably related to different
physical conditions in the parental clouds or different internal
structure for the gas associated with a multi-scale scenario.
Finally, we want to mention that besides the different geometry
and kinematics shown by the GB and the LGD
\citep[see][and references therein]{Eli06a,Eli06b}
the internal distribution of stars with spectral types earlier than
B4 (quantified through the fractal dimension) allows us clearly to
differentiate both systems. This difference could be explained in
terms of a hierarchical star formation scenario.  

\section{CONCLUSIONS}
\label{sec_conclusions}

We have developed an algorithm which uses equations~(\ref{eqnr}) and
(\ref{eqcr}) to estimate the correlation dimension of the stellar
component of the GB and the LGD. The novelty of the algorithm lies
in the implementation of objective criteria to avoid boundary effects
and finite data problems at small scales. We find that early OB
stars (of spectral types earlier than B4) in the GB have a fractal
dimension $\sim 2.7$. This values is very similar to that of the gas
clouds in our Galaxy. This result seems to indicate that younger OB
stars in the GB ``preserve the memory" of the spatial structure of
the cloud where they were born. On the contrary, stars in the GB of
later spectral types show a larger fractal dimension, very similar
to the value found for stars of both age groups in the LGD ($\sim
2.8-2.9$). Several factors (or a combination of them) can contribute
to these morphological differences: age differences among the samples,
different environmental conditions in the birth places, or different
internal structure of the gas in the parental clouds.

\acknowledgments
We want to thank the referee for his/her valuable
comments and criticisms that improved this paper.
E.~J.~A. acknowledges funding from MEC of Spain
through grants AYA2004-05395 and AYA2004-08260-C03-02, and
from Consejer\'{\i}a de Educaci\'on y Ciencia (Junta de
Andaluc\'{\i}a) through TIC-101.

\clearpage

\begin{deluxetable}{ccccc}
\tablecolumns{5}
\tablewidth{0pt}
\tablecaption{The fractal dimension for random point distributions.
The results are the average of ten random realizations. Here we show
(left to right columns) the number of points in each test distribution
($N$), the correlation dimension ($D_c$) with its standard deviation
($\sigma$) calculated both in the three-dimensional space (3D) and in
the projected two-dimensional space (2D).\label{table1}}
\startdata
\tableline
$N$ & $D_c ({\rm 3D})$ & $\sigma ({\rm 3D})$ &
      $D_c ({\rm 2D})$ & $\sigma ({\rm 2D})$ \\
\tableline
2000 & 3.04 & 0.10 & 2.00 & 0.01 \\
1000 & 3.05 & 0.27 & 2.00 & 0.03 \\
 500 & 3.13 & 0.59 & 1.98 & 0.04 \\
 200 & 3.12 & 1.32 & 2.08 & 0.14 \\
\enddata
\end{deluxetable}

\clearpage

\clearpage

\clearpage

\clearpage

\clearpage

\clearpage

\clearpage

\begin{thebibliography}{}

\bibitem[Bergin \& Tafalla(2007)]{Ber07}
Bergin, E. A., \& Tafalla, M.\ 2007, \araa, 45, in press, arXiv:0705.3765
\bibitem[Cartwright \& Whitworth(2004)]{Car04}
Cartwright, A., \& Whitworth, A. P. 2004, \mnras, 348, 589
\bibitem[Cartwright et al.(2006)]{Car06}
Cartwright, A., Whitworth, A. P., \& Nutter, D. 2006, \mnras, 369, 1411
\bibitem[Chappell \& Scalo(2001)]{Cha01}
Chappell, D., \& Scalo, J. 2001, \apj, 551, 712
\bibitem[Ciccotti \& Mulargia(2002)]{Cic02}
Ciccotti, M., \& Mulargia, F. 2002, \pre, 65, 37201
\bibitem[de la Fuente Marcos \& de la Fuente Marcos(2006a)]{Fue06a}
de la Fuente Marcos, R., \& de la Fuente Marcos,  C. 2006, \aap, 452, 163
\bibitem[de la Fuente Marcos \& de la Fuente Marcos(2006b)]{Fue06b}
de la Fuente Marcos, R., \& de la Fuente Marcos,  C. 2006, \aap, 454, 473
\bibitem[de la Fuente Marcos \& de la Fuente Marcos(2006c)]{Fue06c}
de la Fuente Marcos, R., \& de la Fuente Marcos,  C. 2006, \mnras, 372, 279
\bibitem[de Zeeuw et al.(1999)]{Zee99} de Zeeuw, P.~T., Hoogerwerf, R.,
de Bruijne, J.~H.~J., Brown, A.~G.~A., \& Blaauw, A.\ 1999, \aj, 117, 354
\bibitem[Eddy(1977)]{Edd77}
Eddy, W. F. 1977, {\it Trans. Math. Soft.}, 3, 398
\bibitem[Efremov(1995)]{Efr95}
Efremov, Y. N. 1995, \aj, 110, 2757
\bibitem[Elias et al.(2006a)]{Eli06a}
Elias, F., Cabrera-Ca\~no, J., \& Alfaro, E. J. 2006, \aj, 131, 2700
\bibitem[Elias et al.(2006b)]{Eli06b}
Elias, F., Alfaro, E. J., \& Cabrera-Ca\~no, J. 2006, \aj, 132, 1052
\bibitem[Elmegreen et al.(2000)]{Elm00}
Elmegreen, B. G., Efremov, Y., Pudritz, R. E., \& Zinnecker, H. 2000,
in {\it Protostars and Planets IV}, eds. V. Mannings, A. P. Boss, \&
S. S. Russell (Tucson: University of Arizona Press)
\bibitem[Elmegreen \& Elmegreen (2001)]{Elm01}
Elmegreen, B. G., \& Elmegreen, D. M. 2001, \aj, 121, 1507
\bibitem[Elmegreen \& Scalo(2004)]{Elm04}
Elmegreen, B. G., \& Scalo, J. 2004, \araa, 42, 211
\bibitem[Elmegreen(2006)]{Elm06pre} Elmegreen, B. G. 2006, in ASP Conf.
Ser.: {\it Mass loss from stars and the evolution of stellar clusters},
eds. A. de Koter, L. Smith, \& R. Waters (San Francisco: ASP),
astro-ph/0610679
\bibitem[Elmegreen et al.(2006)]{Elm06}
Elmegreen, B. G., Elmegreen, D. M., Chandar, R.,
Whitmore, B., \& Regan, M. 2006, \apj, 644, 879
\bibitem[Falconer(1990)]{Fal90}
Falconer, K. J. 1990, Fractal Geometry: Mathematical
Foundations and Applications (London: Wiley)
\bibitem[Feitzinger \& Galinski(1987)]{Fei87}
Feitzinger, J. V., \& Galinski, T. 1987, \aap, 179, 249
\bibitem[Gould(1879)]{Gou79}
Gould, B. A. 1879, Uranometria Argentina (Buenos Aires: Coni)
\bibitem[Grassberger \& Procaccia(1983)]{Gra83}
Grassberger, P., \& Procaccia, I. 1983, \prl, 50, 346
\bibitem[Guillout et al.(1998)]{Gui98}
Guillout, P., Sterzik, M. F., Schmitt, J. H. M. M.,
Motch, C., \& Neuhauser, R. 1998, \aap, 337, 113
\bibitem[Herschel(1847)]{Her47}
Herschel, J. F. W. 1847, Results of Astronomical Observations made during
the years 1834-1838 at the Cape of Good Hope (London: Smith, Elder, \& Co.)
\bibitem[Khalil et al.(2004)]{Kha04}
Khalil, A., Joncas, G., \& Nekka, F. 2004, \apj, 601, 352
\bibitem[Kitoh et al.(2000)]{Kit00}
Kitoh, S., Kimura, M., Mori, T., Takezawa,
K., \& Endo, S. 2000, {\it Phys. D}, 141, 171
\bibitem[Lindblad et al.(1973)]{Lin73}
Lindblad, P. O., Grape, K., Sanqvist, A., \& Schober, J. 1973, \aj, 24, 309
\bibitem[Mamajek \& Feigelson(2001)]{Mam01} Mamajek, E.~E., \& Feigelson,
E.~D.\ 2001, in ASP Conf. Ser. 244: {\it Young Stars Near Earth: Progress
and Prospects}, eds. R. Jayawardhana \& T. Greene (San Francisco: ASP), 104
\bibitem[Mandelbrot(1983)]{Man83}
Mandelbrot, B. B. 1983,  The Fractal Geometry of Nature, (New York: Freeman)
\bibitem[Moreno et al.(1999)]{Mor99}
Moreno, E., Alfaro,  E. J., \& Franco, J. 1999, \apj, 552, 276
\bibitem[Odekon(2006)]{Ode06}
Odekon, M. C. 2006, \aj, 132, 1834
\bibitem[Parodi \& Binggeli(2003)]{Par03}
Parodi, B. R., \& Binggeli, B. 2003, \aap, 398, 501
\bibitem[Perrot \& Grenier(2003)]{Per03}
Perrot, C. A., \& Grenier, I. A. 2003, \aap, 404, 519
\bibitem[Pisarenko \& Pisarenko(1997)]{Pis95}
Pisarenko, D. V., \& Pisarenko, V. F. 1997, {\it Phys. Lett. A}, 197, 31
\bibitem[S\'anchez et al.(2005)]{San05}
S\'anchez, N., Alfaro, E. J., \& P\'erez, E. 2005, \apj, 625, 849
\bibitem[S\'anchez et al.(2007)]{San07}
S\'anchez, N., Alfaro, E. J., \& P\'erez, E. 2007, \apj, 656, 222
\bibitem[Schmeja \& Klessen(2006)]{Sch06}
Schmeja, S., \& Klessen, R. S. 2006, \aap, 449, 151
\bibitem[Smith(1988)]{Smi88}
Smith, L. A. 1988, {\it Phys. Lett. A}, 133, 283
\bibitem[Stothers \& Frogel(1974)]{Sto74}
Stothers, R., \& Frogel, J. A. 1974, \aj, 79, 456
\bibitem[Torra et al.(2000)]{Tor00}
Torra, J., Fernandez, D., \& Figueras, F. 2000, \aap, 359, 82
\bibitem[Westpfahl et al.(1999)]{Wes99}
Westpfahl, D. J., Coleman, P. H., Alexander,
J., \& Tongue, T. 1999, \aj, 117, 868
\end{thebibliography}
\end{document}